# Debris cloud of India Anti-Satellite Test to Microsat-R Satellite

Yu Jiang

State Key Laboratory of Astronautic Dynamics, Xi'an Satellte Control Center, Xi'an, China

email: jiangyu_xian_china@163.com

**Abstract**

Understanding the motion of debris cloud produced by the anti-satellite test can help us to know the danger of these tests. This study presents the orbit status of 57 fragments observed by the CelesTrak and presented in the NORAD Two-Line Element Sets of India Anti-Satellite Test. There are 10 of these observed fragments have altitudes of the apogee larger than 1000.0km, the maximum one is 1725.7km. We also numerical calculated the number of debris, the results show that the number of debris with the diameter larger than 0.2m is 14, the number of debris with the diameter larger than 0.01m is 6587, and the number of debris with the diameter larger than 0.001m is $7.22 \times 10^5$. The results of the secondary collision of the debris will produced more fragments in the space. The life time of the fragments depends on the initial orbit parameters and the sizes of the debris.

**Keywords**: debris impact; debris cloud; space debris; debris evolution

# 1. INTRODUCTION

The test of Indian anti-satellite weapon, Mission Shakti on Mar. 27, 2019 created a cloud of high-velocity debris in low-Earth orbit (Martin 2019; Hussain and Ahmed 2019; Wolverton 2019). The target was Microsat-R, which was lanched in January, 2019 (Hussain and Ahmed 2019; Wolverton 2019; Arif 2019). The mass of Microsat-R is 740kg, and the orbit is 268km×289km with the inclination 96.60°(Microsat-R 2019). NASA strongly criticised India's anti-satellite weapon test, and pointing out that the debris from the test endangers the International Space Station(ISS)(Ali Khan and Imam 2019; Arif 2019; Martin 2019; Hussain and Ahmed 2019; Johnson-Freese and Burbach 2019; Wolverton 2019); the risk of collision of the debris with the ISS has increased by 44% after the test, NASA tracked about 60 debris objects larger than 10 cm.

The fragments produced by Indian anti-satellite weapon may impact to other debris fragments or spacecrafts. The impact of debris fragments to spacecrafts can cause a sudden permanent local damage of spacecrafts, which can also cause an impulsive orbit change and the attitude deviation of spacecrafts (Lanouette et al. 2015; Krag et al. 2017; Pardini and Anselmo 2017). The impact of debris fragments to spacecrafts/fragments can produce more smaller fragments (Sorge et al. 2016). These collision fragments follow a runaway exponential growth (Rossi et al. 1994; Letizia et al. 2015), which will cause greater risk of collision to spacecrafts. Extension of debris cloud will leads new collision risk with spacecrafts in different orbit planes (Pardini and Anselmo 2011; Letizia 2018; Schaus et al. 2019).

In this paper, we report the result of our calculation of the debris cloud of India anti-satellite test. Our calculation shows that the number of debris fragments larger than 1cm is about 6587, and larger than 0.1 cm is $7.22 \times 10^5$. The impact of other debris with the debris fragments created by India anti-satellite test leads to the secondary ejecta of debris, which can create more debris fragments with more wide distribution in the space.

## 2.METHODS

The equations of motion for debris can be expressed in the inertia space as

$$\ddot{\mathbf{r}} = \mathbf{f}_E + \mathbf{f}_S + \mathbf{f}_M + \mathbf{f}_A + \mathbf{f}_{SR} + \mathbf{f}_L + \mathbf{f}_{PR} \tag{1}$$

where $\mathbf{r}$ is the position vector from the Earth centre to the debris relative to the inertia space. $\mathbf{f}_E = -\nabla V$, $\mathbf{f}_S$ and $\mathbf{f}_M$ are the gravitational acceleration acting on the debris caused by the Earth, Sun, and Moon. $V$ is the potential of the Earth. $\mathbf{f}_A$ and $\mathbf{f}_{SR}$ are accelerations from the atmospheric drag and the solar radiation pressure, respectively. $\mathbf{f}_L$ is the acceleration from the Lorentz force of the Earth magnetic field, $\mathbf{f}_{PR}$ is the acceleration from the Poynting-Robertson drag.

The gravitational potential (Cunningham 1970) can be calculated by

$$V = \frac{GM_E}{r} \sum_{n=0}^{N} \sum_{m=0}^{n} \left(\frac{R_e}{r}\right)^n \left(\bar{C}_{nm} \cos m\lambda + \bar{S}_{nm} \sin m\lambda\right) \bar{P}_{nm} \sin(\psi). \tag{2}$$

Here G is the gravitational constant, $M_E$ is the Earth mass, $r = |\mathbf{r}|$, $R_e$ is the Earth equatorial radius. $\lambda$ and $\psi$ are longitude and latitude, respectively. $\bar{C}_{nm}$ and $\bar{S}_{nm}$ are normalized spherical harmonics coefficients of order $n$ and degree $m$, $\bar{P}_{nm}$ is the normalized Legendre polynomials, $N$ is the maximum order the model used.

The acceleration of the Solar gravity and the Moon gravity are

$$\mathbf{f}_S = -GM_S \left(\frac{\mathbf{r}_{DS}}{r_{DS}^3} - \frac{\mathbf{r}_S}{r_S^3}\right), \tag{3}$$

$$\mathbf{f}_M = -GM_M \left( \frac{\mathbf{r}_{DM}}{r_{DM}^3} - \frac{\mathbf{r}_M}{r_M^3} \right). \tag{4}$$

Here $M_S$ and $M_M$ are the mass of the Sun and Moon, respectively. $\mathbf{r}_{DS}$ and $\mathbf{r}_{DM}$ are position vectors from the Sun/Moon to the debris, respectively. $\mathbf{r}_S$ and $\mathbf{r}_M$ are position vectors from the Earth centre to the Sun and Moon, respectively. $r_{DS}$ and $r_{DM}$ are norms of $\mathbf{r}_{DS}$ and $\mathbf{r}_{DM}$, respectively. $r_S = |\mathbf{r}_S|$, $r_M = |\mathbf{r}_M|$.

The acceleration from the solar radiation pressure (Burns et al. 1979) is

$$\mathbf{f}_{SR} = \left( \frac{S_0 A_s}{m_d c} \right) \left( \frac{r_{s0}}{r_{sd}} \right)^2 Q_{pr} \mathbf{r}_{SR}, \tag{5}$$

where the solar constant $S_0 = 1.3608 \times 10^6$ ergs·cm$^{-2}$·s$^{-1}$, $A_s$ is the cross sectional area of the debris related to the radiation, $m_d$ is the debris mass, $c$ is the light velocity, $r_{s0} = 1\text{AU}$, $r_{sd}$ is the distance from the sun to the debris, $Q_{pr}$ is the radiation pressure coefficient, $\mathbf{r}_{SR}$ is the direction from the sun to the debris.

The acceleration from the atmospheric drag is

$$\mathbf{f}_a = -\frac{1}{2} \frac{C_d A}{m_d} \rho v \mathbf{v} \tag{6}$$

Where the drag coefficient $C_d = 2.2$, $A$ is the cross sectional area of the debris related to the drag, $\rho$ is the atmospheric density, $\mathbf{v}$ is the velocity of the debris in the inertia space, $v = |\mathbf{v}|$. The atmospheric density here is computed using the NRLMSISE‐00 empirical atmospheric model (Picone et al. 2002).

The geomagnetic field is

$$\mathbf{B}(r, \theta, \phi, t) = -\nabla V_{mag}, \tag{7}$$

the magnetic scalar potential $V_{mag}$ (Thébault et al. 2015) is

$$V_{mag} = a \sum_{n=0}^{N_{gm}} \sum_{m=0}^{n} \left(\frac{a}{r}\right)^{n+1} \left(g_n^m(t)\cos(m\phi) + h_n^m(t)\sin(m\phi)P_n^m(\cos\theta)\right). \tag{8}$$

Here the mean geomagnetic reference spherical radius $a=6,371.2$ km. $\theta$ and $\varphi$ are latitude and east longitude, respectively. $t$ is the time. $n$ and $m$ are degree and order, respectively. $g_n^m$ and $h_n^m$ are geomagnetic Gauss coefficients, and $P_n^m(\cos\theta)$ is the normalized Legendre functions. $N_{gm}$ is the maximum order the geomagnetic model used.

The acceleration from the Lorentz force is

$$\mathbf{F}_L = \frac{4\pi\varepsilon_0 Us}{m_d}(\dot{\mathbf{r}} - \boldsymbol{\omega} \times \mathbf{r}) \times \mathbf{B}(r,\theta,\phi,t), \tag{9}$$

where the permittivity $\varepsilon_0 = 8.854187817 \times 10^{12} F \cdot m^{-1}$, the surface potential of the debris $U = +5V$ (Grün et al. 1994), $s$ is the debris radius, $\boldsymbol{\omega}$ is the angular velocity vector of the Earth.

The acceleration from the Poynting-Robertson drag (Burns et al. 1979) is

$$\mathbf{F}_{PR} = -\left(\frac{S_0 A}{m_d c^2}\right)\left(\frac{r_{s0}}{r_{sd}}\right)^2 Q_{pr}\dot{\mathbf{r}}_{DS}, \tag{10}$$

The size distribution of debris can be expressed in mass using the power-law equation (Johnson et al .2001; Liou et al. 2002; Sakuraba et al. 2008)

$$N_{debris}(\geq M_f) = \alpha \left(\frac{M_f}{M_{tot}}\right)^{-0.68}. \tag{11}$$

Where $N_{debris}$ is the number of debris weight larger than $M_f$. $M_{tot}$ is the total mass of debris, $\alpha$ is the correction factor which is set to be 0.78 in Sakuraba et al. (2008). The breakup model may have large error relative to the real test (Liu et al. 2012). Thus we use the real measure to determine the correction factor. Main materials of the satellite is assumed to be aluminum alloy and have the density $\rho = 2.7 \times 10^3 kg \cdot m^{-3}$. Then the power-law equation can be written as

$$N_{debris}(\geq d) = \alpha \left(\frac{\pi \rho d^3}{6M_{tot}}\right)^{-0.68} = \beta \left(\frac{\rho d^3}{M_{tot}}\right)^{-0.68} \quad (12)$$

Where $\beta$ is the new correction factor. $d$ is the diameter of the debris.

For Microsat-R, $M_{tot} = 740kg$. If we use $\alpha = 0.78$, then the number of debris which is larger than 0.1m is 55, i.e. $N_{debris}(\geq 0.1m) = 55$, $N_{debris}(\geq 0.01m) = 6038$, $N_{debris}(\geq 0.001m) = 6.62 \times 10^5$. NASA reported that the number of debris fragments which is larger than 0.1m is about 60 (Gill 2019). Thus the new correction factor is $\beta = 1.32$ for India's antisatellite test in 2019. With this new factor, the results are: $N_{debris}(\geq 0.2m) = 14$, $N_{debris}(\geq 0.1m) = 60$, $N_{debris}(\geq 0.01m) = 6587$, $N_{debris}(\geq 0.0018m) = 2.18 \times 10^5$, $N_{debris}(\geq 0.001m) = 7.22 \times 10^5$.

The distribution density function (Krivov et al. 2003; Szalay and Horányi 2016) of the velocity for debris relative to Microst-R before impact is calculated by

$$f(\hat{v}) = \frac{\delta \hat{v}}{(1-\hat{v}^2)^2} e^{-\frac{\beta \hat{v}}{1-\hat{v}^2}}, \quad (13)$$

Where $\hat{v} = v_{Col}/v_m$, $v_{Col}$ is the collision velocity of the impactor, $v_m$ is the maximum ejection velocity of debris, $\beta = 8.69, \delta = 7.2 \times 10^{-3} s \cdot m^{-1}$.

## 3.RESULTS

**Observed debris of India anti-satellite test and calculation.** Data of two line element had been downloaded from web on Apr. 5, 2019 (NORAD 2019). Orbits are calculated using SGP4 on UTC time 0:00, Apr. 5, 2019. There are 57 fragments in the file of two line element for Indian ASAT Test Debris (See Fig.1). Among these 57 fragments, 10 of them have altitudes of the apogee larger than 1000.0km (See Fig.2 a). The maximum of the altitude of the apogee is 1725.7km. The range of the altitudes of the apogee of the 57 fragments is 248km to 1725.7km, and the range of the altitudes of the perigee of them is 208km to 288.6km. The maximum of the eccentricity is 0.0952 (See Fig.2 b). The range of the inclination is 94.9 deg to 96.8 deg (See Fig.2 c,d,e). The debris fragments may impact to low earth orbit satellites and the International Space Station. The impact between two different spacecrafts, or debris to spacecraft, may cause huge accident in the space, such as the impact of Cosmos 2251 and Iridium 33 in 2009 generated a large number of debris (Anselmo and Pardini 2009). Our calculation shows that the number of debris fragments larger than 0.2m is 14, larger than 0.01m is 6587, larger than 0.0018m is $2.18 \times 10^5$, and larger than 0.001m is $7.22 \times 10^5$.

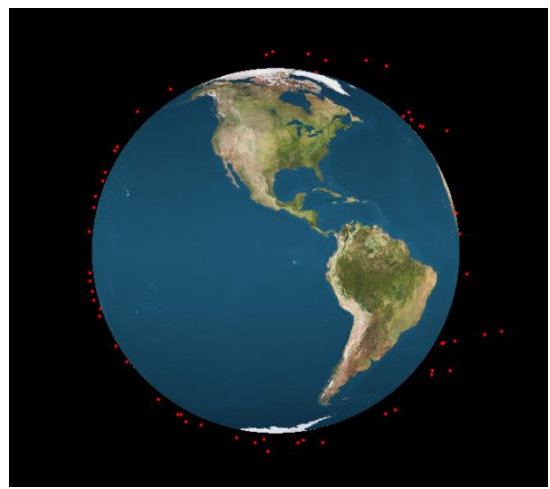

Fig. 1| Distribution of debris cloud of Microsat-R

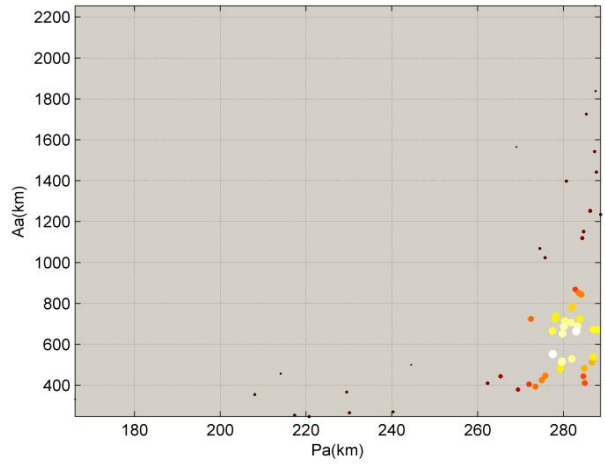

a

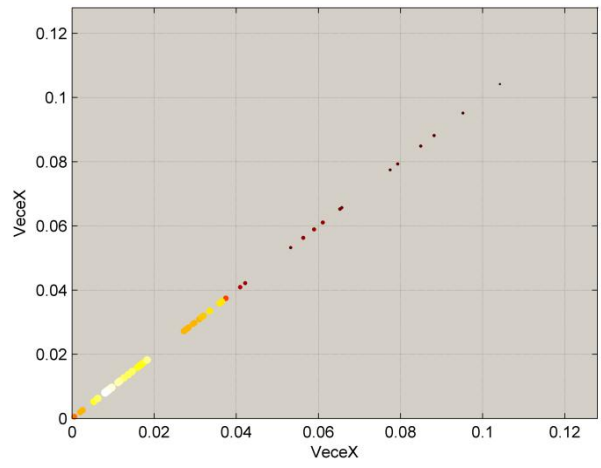

b

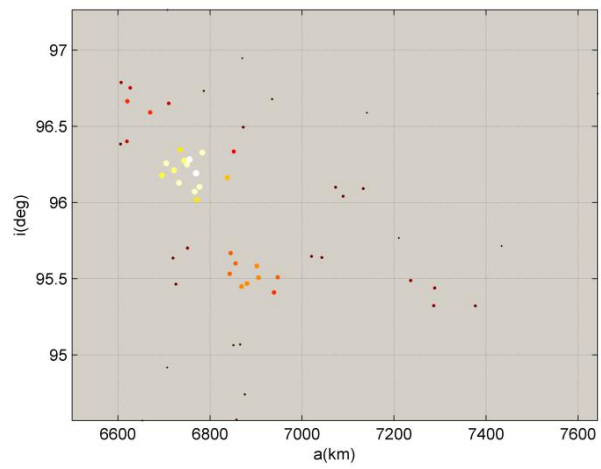

c

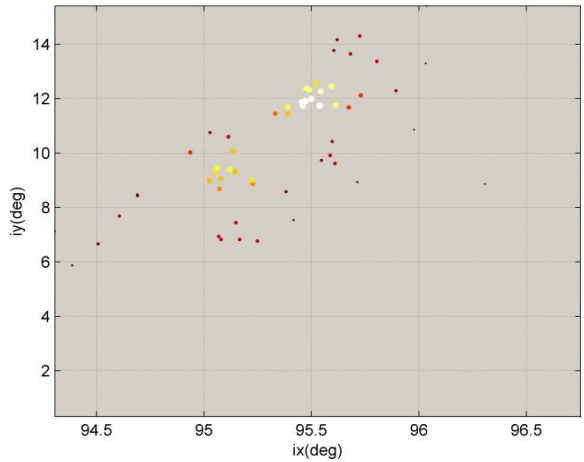

d

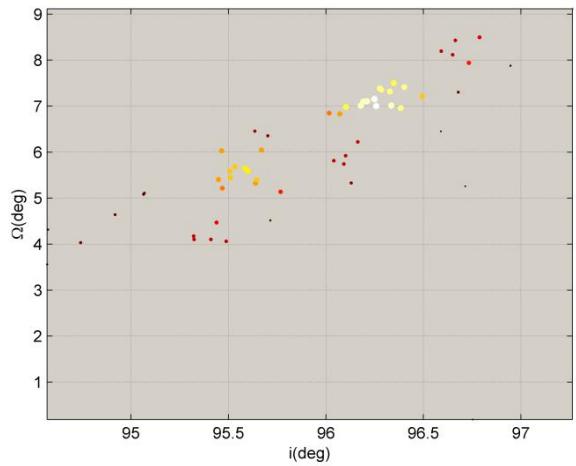

e

Fig. 2| Debris in the plane of orbital parameters. **a**, altitude of perigee and altitude of the apogee. The Earth radius used here is 6371km. **b**, eccentricity vector. **c**, semi-major axis and inclination. **d**, inclination vector. **e**, inclination and right ascension of ascending node.

**Secondary ejecta of debris produced by India anti-satellite test.** The secondary ejecta of other debris impacts a spacecraft/debris surface may also produce debris (Schonberg 2001; Mandeville and Bariteau 2004). To understand the contribution of secondary ejecta to the total debris population in the space produced by the fragments from Indian anti-satellite weapon and fragments previously existed, we consider the possible secondary ejecta of the fragments produced by Indian anti-satellite weapon. The size and velocity distribution of secondary debris fragments are modeled using the distribution density function. We calculated the secondary ejecta of debris clouds from Object C of the Data file of two line element of Indian anti-satellite weapon (see Table 1). We modeled the motion of 500 debris fragments (see Fig. 3). Among these 500 fragments, 296 fragments entry 50.0km height relative to the earth surface. 130 fragments have the altitudes of perigee larger than 200.0km (see Fig. 4). Among these 130 fragments, 47 fragments have the altitudes of apogee larger than 1000.0km. The trajectories of secondary debris clouds show that for the fragment trajectories with the apogee height larger than the height of secondary ejecta, the perigee are at the positions of secondary ejecta, and for the fragment trajectories with the perigee height smaller than the height of secondary ejecta, the apogee are at the positions of secondary ejecta. For the debris fragments, the interval of the semi-major axis is [5038.6, 11255.0] km, the interval of the eccentricity is [0.00165, 0.41398]. The distribution of the eccentricity vector is dense when the eccentricity vector is near the origin point [0.0, 0.0], and sparse when it is far from the origin point. The interval of the inclination and the right ascension of ascending node are [92.589, 100.729] deg and [25.861, 40.660] deg, respectively. The scatter diagram of the inclination vector looks like a slash with the center part denser, while the scatter diagram of the eccentricity vector looks like a disc with the center part denser.

Table 1 The orbital parameters of Object C of Indian anti-satellite weapon (Earth radius is 6371km)

| | |
|---|---|
| **semi-major axis** | **7286.1874km** |
| **eccentricity** | 0.0861779 |
| **inclination** | 95.3237 deg |
| **right ascension of ascending node** | 4.1062 deg |
| **argument of perigee** | 350.0688deg |
| **mean anomaly** | 8.4347deg |
| **orbital period** | 103.1598min |
| **altitude of the perigee** | 287.279km |
| **altitude of the apogee** | 1543.096km |

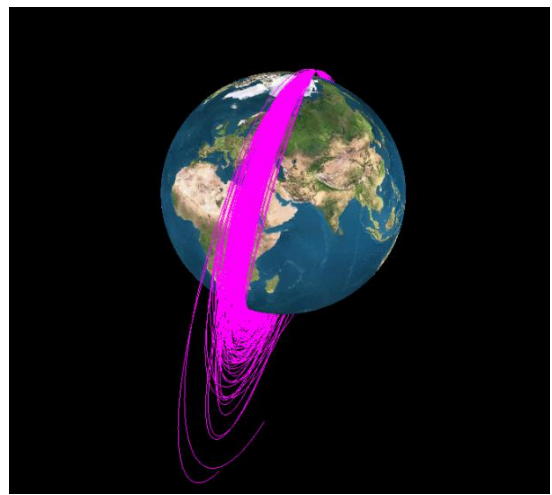

Fig. 3| Trajectories of secondary debris clouds. The trajectories are plotted in the inertia space.

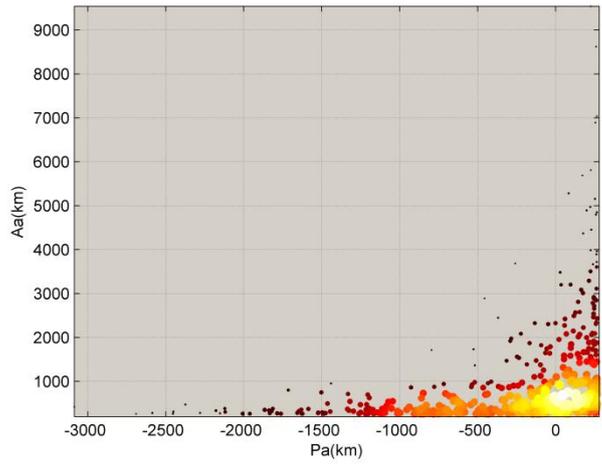

a

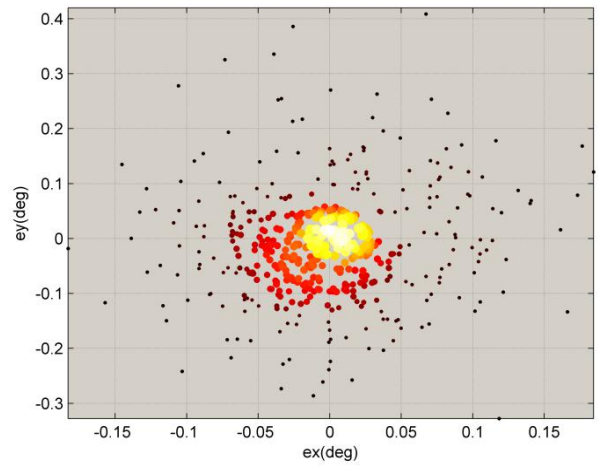

b

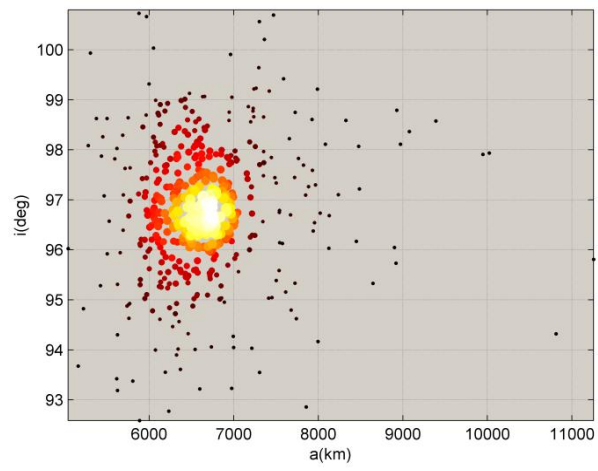

c

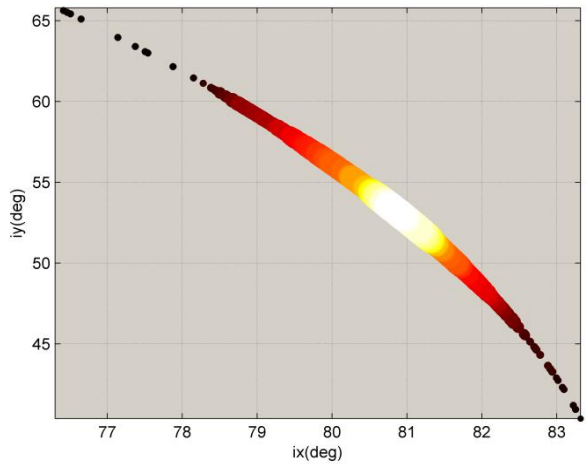

d

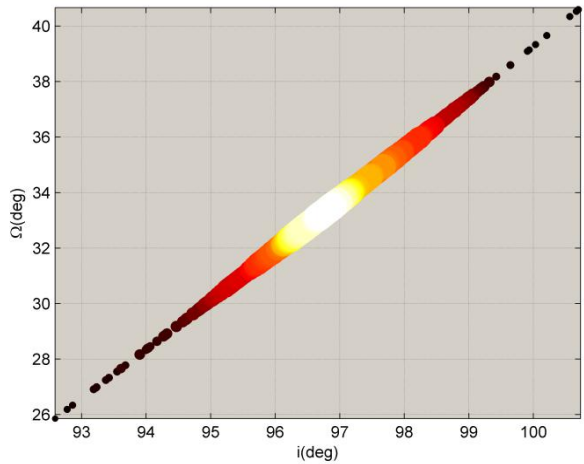

e

Fig. 4| Secondary debris in the plane of orbital parameters. **a**, altitude of perigee and altitude of the apogee. The Earth radius used here is 6371km. **b**, eccentricity vector. **c**, semi-major axis and inclination. **d**, inclination vector. **e**, inclination and right ascension of ascending node.

## 4.DISSCUSSION

If the sizes of debris fragments become smaller, the number of debris fragments becomes larger. Because when the size of a fragment is small, the area-mass ratio of the fragment becomes large, and the atmospheric drag becomes large, then the small fragments may entry atmosphere more quickly. Big fragments are more dangerous when impact on to spacecrafts. However, steel sphere with the size of 1.8 mm diameter and relative speed of 2.5km·s$^{-1}$ can penetrate steel target which is 2.67mm thick (Lamberson and Rosakis 2013). Compare this with our results: the number of debris fragments larger than 0.0018m is $2.18\times 10^5$, we know that the fragments with potential danger produced by India anti-satellite test have the number larger than $2.18\times 10^5$.

We consider the potential danger for orbital regions. The maximum value of the altitude of the apogee of observed debris: 1725.7km, this is near the maximum height of Low Earth Orbit (LEO). This implies the produced debris of India anti-satellite test have dangerous for most of spacecrafts in LEO. In addition, the secondary ejecta of debris fragments produced by India anti-satellite test can also generate lots of new fragments. This will further increase the possible impact danger.